\documentclass[3p,onecolumn,times]{elsarticle}
\usepackage[usenames]{color}
\usepackage{graphicx}
\usepackage{amssymb}
\usepackage{amsmath}

\journal{Communications in Nonlinear Science and Numerical Simulation}

\begin{document}
\biboptions{sort&compress}

\begin{frontmatter}

\title{Bright soliton dynamics in  Spin Orbit-Rabi coupled Bose-Einstein condensates}

\author[kmu,uae]{P.~S.~Vinayagam}
\author[kmu]{R.~Radha\corref{cor1}}
\ead{vittal.cnls@gmail.com}
\cortext[cor1]{Corresponding author}
\address[kmu]{Centre for Nonlinear Science (CeNSc), PG and Research Department of Physics, Government College for Women (Autonomous), Kumbakonam 612001, India}
\address[uae]{Department of Physics,
United Arab Emirates University, P.O.Box 15551, Al-Ain, United
Arab Emirates}
\author[bdu]{S.~Bhuvaneswari}
\author[bdu]{R. Ravisankar}
\author[bdu]{P. Muruganandam}
\ead{anand@cnld.bdu.ac.in}
\address[bdu]{Department of Physics, Bharathidasan University,  Palkaliperur Campus, Tiruchirapalli 620024, India}

\begin{abstract}
We investigate the dynamics of a spin-orbit (SO) coupled  BECs in
a time dependent harmonic trap and show the dynamical system to be
completely integrable by constructing the Lax pair. We then employ
gauge transformation approach to witness  the rapid oscillations
of the condensates for a relatively smaller value of SO coupling
in a time independent harmonic trap compared to their counterparts
in a transient trap. Keeping track of the evolution of the
condensates in a transient trap during its transition from
confining to expulsive trap, we notice that they collapse in the
expulsive trap. We further show that one can manipulate the
scattering length through Feshbach resonance to stretch the
lifetime of the confining trap and revive the condensate.
Considering a SO coupled state as the initial state, the numerical
simulation indicates that the reinforcement of Rabi coupling on SO
coupled BECs generates the striped phase of the bright solitons
and does not impact the stability of the condensates despite
destroying the integrability of the dynamical system.
\end{abstract}


\end{frontmatter}

\section{Introduction}
The advent of Bose-Einstein condensates (BECs) in
rubidium~\cite{ref3} and the subsequent experimental
identification of bright~\cite{bright,bright1,bright2} and dark
solitons~\cite{dark,dark1} for attractive and repulsive binary
interaction, respectively contributed to a resurgence in the
investigation of ultra cold matter. At ultralow temperatures, the
macroscopic wave function of BECs can be described by the mean
field Gross-Pitaevskii (GP) equation, which is essentially a
variant of the celebrated nonlinear Schr\"odinger (NLS)
equation~\cite{meanfield}. It is worth pointing out at this
juncture that the behavior of single (scalar) component BECs is
influenced by the external trapping potential and binary
interatomic interaction. Experimental realization of vector (or
two component) BECs in which two (or more) internal states or
different atoms can be populated has given a fillip to the
investigation of multi-component BECs. In contrast to the single
component BECs, the multi-component BECs exhibit rich dynamics by
virtue of inter-species and intra species binary interaction which
can be either attractive or repulsive. This extra freedom
associated with multi-component BECs enables them to display novel
and rich phenomenon like multidomain
walls~\cite{domainwalls,domainwalls1,domainwalls2}, spin switching
soliton pairs (either bright-bright~\cite{b-b,b-b1,b-b2},
dark-dark~\cite{d-d} or bright-dark, etc) which can never be
witnessed in single component BECs.

Recently, in a landmark experiment, Spielman group at NIST have
engineered a  synthetic spin-orbit (SO) coupling for a BEC
\cite{SOC-experiment}. In the experiment, two Raman laser beams
were used to couple a two component BEC consisting of
(predominantly) two hyperfine states of $^{87}$Rb. The momentum
transfer between laser beams and atoms contributes to the rich
possibility of creating synthetic electric and magnetic fields. Recent investigations have explored the possibility of identifying tunable spin orbit coupled BECs~\cite{tunable,two-dim,vortex,loc,bichromatic-soc,doublewell} with various trapping potentials and stable regimes of condensates have been observed. The identification of stripe phase in the investigation of spin orbit coupled BECs \cite{kueisun} has only reignited the enthusiasm in this domain of interest as the striped phase is believed to be an indication of the phase transition taking place in a spin orbit coupled BEC. Also, the influence of spin-orbit coupling in a one-dimensional BEC, particularly the interplay of the SOC, Raman coupling, and nonlinearity induced precession of the soliton's spin has been studied recently~\cite{Wen2016}. The existence of ``Striped phase", which essentially consists of a linear combination of plane waves and this has contributed to the idea of using ultra cold atoms for the implementation of a quantum simulator \cite{quantumsimulator}.

At this juncture, it should be mentioned that eventhough nonlinear
excitations like vortices \cite{vortices},  Skymions
\cite{skymions} and bright solitons \cite{moto} have been
generated in a SO coupled BEC, the dynamics of SO coupled BECs in
a time dependent harmonic trap governed by a two coupled GP
equation has not been exactly solved analytically. In this paper,
we construct the linear eigenvalue problem of SO coupled GP
equation in a transient harmonic trap and show that it is
completely integrable. We then generate bright solitons solutions
and track their evolution in a transient harmonic trap. We observe
that the addition of SO coupling contributes to the rapid
oscillations of real and imaginary parts of the order parameter
and this occurs at a relatively lower value of SO coupling
parameter in a time independent harmonic trap compared to their
counterparts in a transient trap. Tracing the evolution of the
bright solitons (or the condensates) during its transition from
confining to expulsive trap, we notice that the condensates
collapse suddenly in the expulsive trap. By employing Feshbach
resonance management, we show that one can retrieve the
condensates by stretching the lifespan of confining trap. Then
considering a SO coupled state as the initial state, we
numerically study the impact of Rabi coupling on the condensates.
The results of our investigation indicate that the reinforcement
of Rabi coupling on SO coupled BECs generates striped solitons and
leaves no impact on the stability of BECs. We also emphasize that
all the above occurs despite the transition of the dynamical
system to the nonintegrable regime.

\section{The model and Lax pair}

We consider a spin-orbit coupled quasi-one dimensional BEC in a
parabolic trap with longitudinal and transverse frequencies
$\omega_x \ll \omega_\perp$. Assuming equal contributions of
Rashba \cite{rashba} and Dresselhaus \cite{dressel-soc} SO
coupling (as in the experiment of Ref.\cite{SOC-experiment}),
which can be described at sufficiently low temperatures by a set
of coupled GP equation of the form \cite{ho96,esry97,pu98}:
\begin{subequations}\label{eq:gpe}
\begin{align}
\mathrm{i} \frac{\partial \psi_{1}}{\partial t}
   = & \left[-\frac{1}{2}\frac{\partial^2}{\partial x^2}+V(x,t)-\gamma(t)\left(\lvert \psi_{1}\rvert^2+\lvert \psi_{2}\rvert^2\right)
     -\mathrm{i}k_L\frac{\partial}{\partial x} \right] \psi_{1} +\Omega \psi_{2},\label{eq:gpe1a} \\
\mathrm{i}  \frac{\partial \psi_{2}}{\partial t}
   = & \left[-\frac{1}{2}\frac{\partial^2}{\partial x^2}+V(x,t)-\gamma(t)\left(\lvert \psi_{1}\rvert^2+\lvert \psi_{2}\rvert^2\right)
      +\mathrm{i}k_L\frac{\partial}{\partial x} \right] \psi_{2} +\Omega \psi_{1}. \label{eq:gpe1b}
\end{align}
\end{subequations}%
In the above equation, the term $\pm i k_L \partial / \partial x$
represents the momentum transfer between the laser beams and atoms
due to SO coupling, $\gamma(t)$ represents binary attractive
interaction, and the linear cross coupling parameter $\Omega$
denotes Rabi coupling and $V(x,t)= \lambda(t)^2 x^2 /2$, where
$\lambda(t)={\omega_x}/{\omega_\perp}$, is the time dependent trap
frequency

Switching off the Rabi coupling ($\Omega=0$), and employing the
following transformation
\begin{subequations}\label{eq:sotrans}
\begin{align}
\psi_{1}(x,t) & = q_1(x,t) \exp \left[ \frac{\mathrm{i}}{2} k_L \left(k_L t - 2 x \right) \right], \\
\psi_{2}(x,t) & = q_2(x,t) \exp \left[ \frac{\mathrm{i}}{2} k_L \left(k_L t + 2 x \right) \right] ,
\end{align}%
\end{subequations}
Equation~(\ref{eq:gpe}) can be written in a simpler form by
eliminating the SO coupling term as,
\begin{subequations}\label{eq:gpe2}
\begin{align}
\mathrm{i}  \frac{\partial q_{1}}{\partial t}
   = & \left[-\frac{1}{2}\frac{\partial^2}{\partial x^2}+V(x,t)-\gamma(t)\left(\lvert q_{1}\rvert^2+\lvert q_{2}\rvert^2\right) \right] q_1,  \label{eq:gpe2a} \\
\mathrm{i}  \frac{\partial q_{2}}{\partial t}
   = & \left[-\frac{1}{2}\frac{\partial^2}{\partial x^2}+V(x,t)-\gamma(t)\left(\lvert q_{1}\rvert^2+\lvert q_{2}\rvert^2\right) \right] q_2. \label{eq:gpe2b}
\end{align}
\end{subequations}%
We emphasize that the model governed by equations (\ref{eq:gpe})
is \textit{exactly integrable} if either Rabi coupling (Zeeman
splitting) ($\Omega$) or SO coupling ($ i k_L$) is taken into
account, but not both of them \cite{konotop}.

The above coupled equations (\ref{eq:gpe2a}) and (\ref{eq:gpe2b}) admit
the following Lax-pair
\begin{subequations}%
\begin{align}%
\Phi_{x} +{\cal U}\Phi =& 0,  \label{Phix} \\
\Phi_{t} +{\cal V}\Phi =& 0,  \label{Phit}
\end{align}
\end{subequations}%
where $\Phi =(\phi_{1},\phi
_{2},\phi_{3})^{T}$ is a three-component Jost function,
\begin{align}
{\cal U}=\begin{pmatrix}
i\zeta (t) & {\cal U}_{12} & {\cal U}_{13} \\
{\cal U}_{21} & -i\zeta (t) & 0 \\
{\cal U}_{31} & 0 & -i\zeta (t)%
\end{pmatrix},
\;\;\;
{\cal V}=
\begin{pmatrix}
{\cal V}_{11} & {\cal V}_{12} & {\cal V}_{13} \\
{\cal V}_{21} & {\cal V}_{22} & {\cal V}_{23} \\
{\cal V}_{31} & {\cal V}_{32} & {\cal V}_{33}%
\end{pmatrix},
\label{UV}
\end{align}
with
\begin{align}
{\cal U}_{12} = & \sqrt{\gamma (t)}q_{1}(x,t) \exp \left[\mathrm{i} \phi(x,t)\right], \notag \\
{\cal U}_{13} = & \sqrt{\gamma (t)}q_{2}(x,t) \exp \left[\mathrm{i} \phi(x,t)\right], \notag \\
{\cal U}_{21} = & -\sqrt{\gamma (t)}q_{1}^{\ast}(x,t) \exp \left[ -\mathrm{i} \phi(x,t)  \right], \notag \\
{\cal U}_{31} = & -\sqrt{\gamma (t)}q_{2}^{\ast}(x,t) \exp \left[ -\mathrm{i} \phi(x,t)  \right], \\
{\cal V}_{11} = &\mathrm{i} \zeta(t)\left[ c(t)x  - \zeta (t) \right]
          +\frac{\mathrm{i}}{2}\gamma(t) \left( \lvert q_{1}(x,t)\rvert^2+\lvert q_{2}(x,t)\rvert^2 \right), \notag \\
{\cal V}_{12} = &\sqrt{\gamma (t)}  \left[c(t)x-\zeta (t)\right] q_{1}(x,t) \exp  \left[ \mathrm{i}\phi (x,t) \right] 
          +\frac{\mathrm{i}}{2}\sqrt{\gamma (t)}\big[q_{1}(x,t) \exp  \left[ \mathrm{i}\phi (x,t) \right]\big]_{x},  \notag \\
{\cal V}_{13} = &\sqrt{\gamma (t)}\left[c(t)x-\zeta (t)\right]q_{2}(x,t) \exp \left[ \mathrm{i}\phi(x,t) \right] 
         +\frac{\mathrm{i}}{2}\sqrt{\gamma (t)}\big[q_{2}(x,t) \exp  \left[ \mathrm{i}\phi (x,t) \right]\big]_{x},
\notag \\
{\cal V}_{21} = &-\sqrt{\gamma(t)} \left[c(t)x-\zeta (t)\right] q_{1}^{\ast}(x,t) \exp \left[ -\mathrm{i}\phi (x,t) \right] 
         +\frac{\mathrm{i}}{2}\sqrt{\gamma(t)}\big[q_{1}^{\ast}(x,t)\exp\left[-\mathrm{i}\phi(x,t)\right]\big]_{x},  \notag \\
{\cal V}_{22} = & -\mathrm{i} \zeta(t)\left[ c(t)x  - \zeta (t) \right]
-\frac{\mathrm{i}}{2} \gamma (t) \lvert q_{1}(x,t)\rvert^2 ,  \notag \\
{\cal V}_{23} = & -\frac{\mathrm{i}}{2} \gamma (t)q_{1}^{\ast}(x,t) q_{2}(x,t) ,  \notag \\
{\cal V}_{31} = & -\left[c(t)x-\zeta (t)\right] \sqrt{\gamma (t)} q_{2}^{\ast}(x,t) \exp  \left[ \mathrm{i}\phi (x,t) \right] 
+ \frac{\mathrm{i}}{2}\sqrt{\gamma (t)}\big[q_{2}^{\ast}(x,t) \exp  \left[ \mathrm{i}\phi (x,t) \right]\big]_{x},  \notag \\
{\cal V}_{32} = &-\frac{\mathrm{i}}{2} \gamma (t) q_{1}(x,t)q_{2}^{\ast}(x,t) \exp \left[ 2 \mathrm{i}\phi (x,t)\right] ,  \notag \\
{\cal V}_{33} = &-\mathrm{i} \zeta(t)\left[ c(t)x  - \zeta (t) \right]
-\frac{\mathrm{i}}{2} \gamma (t) \lvert q_{2}(x,t)\rvert^2 .
\end{align}%
In the above, $\phi (x,t) \equiv c(t)x^{2}/2 $, and $\left[ \ldots
\right]_x$ denotes differentiation with respect to $x$. The
compatibility condition $ {\cal U}_{t}-{\cal V}_{x}+[{\cal
U},{\cal V}]=0$ generates the SO coupled GP equation
(\ref{eq:gpe}) without  Rabi coupling ($\Omega=0$), while the
spectral parameter $\zeta (t)$ obeys the following equation:
\begin{align}
\frac{d}{dt}\zeta (t)= c(t)\zeta (t),  \label{zeta-prime}
\end{align}%
with
\begin{align}
\lambda(t)^2= \frac{d}{dt}c(t)-c(t)^{2},
\label{trap}
\end{align}%
and
\begin{align}
c(t) = \frac{d}{dt}\ln \gamma (t).
\label{tf}
\end{align}%
It may be noted that a similar Riccati equation~(\ref{trap}) has
been employed to solve GP-type
equations~\cite{riccati,riccati1,riccati2,riccati3,riccati4}. In
fact, the identification of the Riccati-type equation~(\ref{trap})
gives the first signature of complete integrability of
Equation~(\ref{eq:gpe}) with ($\Omega=0$). Equation~(\ref{trap}),
which determines the parabolic potential strength, $\lambda
^{2}(t)$, demonstrates that it is related to the interaction
strength, $\gamma (t)$, through the integrability condition, which
can be derived by simply substituting Equation~(\ref {tf}) in
Equation~(\ref{trap}):
\begin{align}
\gamma(t) \frac{d^2}{dt^2}\gamma (t) - 2\left( \frac{d}{dt}\gamma (t)\right) ^{2} + \lambda(t)^{2}\gamma ^{2}(t)=0.  \label{integ}
\end{align}
Thus, the system of coupled GP equations~(\ref{eq:gpe2}) is
completely integrable for suitable choices of $\lambda (t)$ and
$\gamma (t)$, which are consistent with equation (\ref{integ}).
For condensates in a time independent harmonic trap, $\lambda (t)=
\lambda_0$ (a constant), Equation~(\ref{integ}) yields $\gamma
(t)=\gamma_0 \exp \left(\lambda_0 t \right)$, where $\gamma_0$ is
an arbitrary constant.

Considering a vacuum seed solution, i.e., $q_1(x,t)=q_2(x,t)=0$,
and employing gauge transformation approach
\cite{llc,gaugetrans,gaugetrans1,gaugetrans2}, we obtain
\begin{subequations}%
\label{sol:q}
\begin{align}
q_{1}(x,t)=2 \varepsilon_{1} \frac{  \beta(t)}{\sqrt{\gamma(t)}} \exp \left[\mathrm{i}\xi(x,t)-\mathrm{i}\phi (x,t) \right] \text{sech\,} \theta(x,t), \label{sol:q1}\\
q_{2}(x,t)=2 \varepsilon_{2} \frac{ \beta(t)}{\sqrt{\gamma(t)}} \exp \left[\mathrm{i}\xi(x,t)-\mathrm{i}\phi (x,t) \right] \text{sech\,} \theta(x,t), \label{sol:q2}
\end{align}%
\end{subequations}%
where
\begin{subequations}%
\label{eq:xiphi}
\begin{align}
\theta(x,t) & = 2 \beta(t) x - 4\int\alpha(t)\beta(t) \,  dt + 2\delta, \\
\xi(x,t)    & = 2 \alpha(t) x  - 2\int \left[ \alpha(t)^{2}-\beta(t)^{2}\right] dt-2\chi, \\
\alpha(t)   & = \alpha_{0} \exp \int c(t)\,dt, \;\;\; \beta(t) = \beta_{0} \exp  \int c(t)\, dt, \\
\gamma(t)   & = \gamma_0 \exp \int c(t)\,dt,  \;\;\;\phi(x,t) =\frac{1}{2}c(t) x^{2},
\end{align}%
\end{subequations}%
$\delta$ and $\chi$ are arbitrary real parameters, and
$\varepsilon_{1}$ and $\varepsilon_{2}$ are coupling constants
subject to the constraint
$\lvert\varepsilon_{1}\rvert^{2}+\lvert\varepsilon_{2}\rvert^{2}=1$.
A formal solution to the coupled GP equation (\ref{eq:gpe}), in
the absence of Rabi term, can be straightforwardly written using
the transformation (\ref{eq:sotrans}). The solutions given by Equations (2) illustrate the momentum transfer between laser and atoms when a vector BEC with order parameters $q_1$ and $q_2$  driven by bright solitons given by Equations (12) is irradiated with a laser beam.

\section{Soliton dynamics of spin-orbit-Rabi coupled BEC in a transient and time independent trap}

\begin{figure}[!ht]
\centering\includegraphics[width=0.6\linewidth]{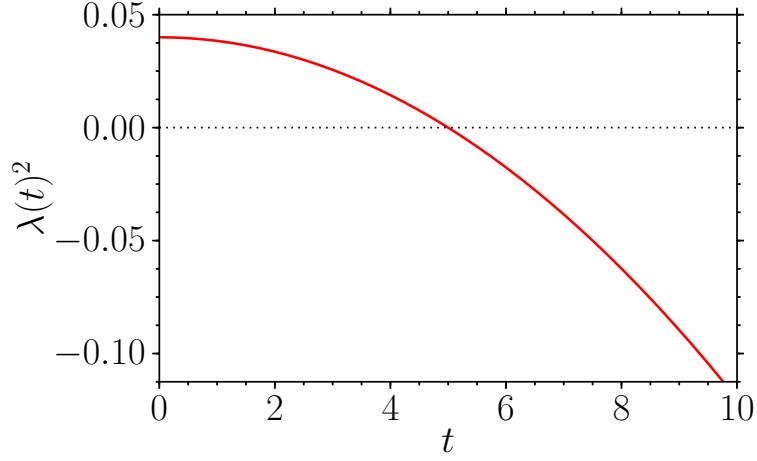}
\caption{Evolution of the time dependent trap for
$c(t)=0.04t$}\label{Fig.1.trap}
\end{figure}

\begin{figure}[!ht]
\centering\includegraphics[width=0.9\linewidth]{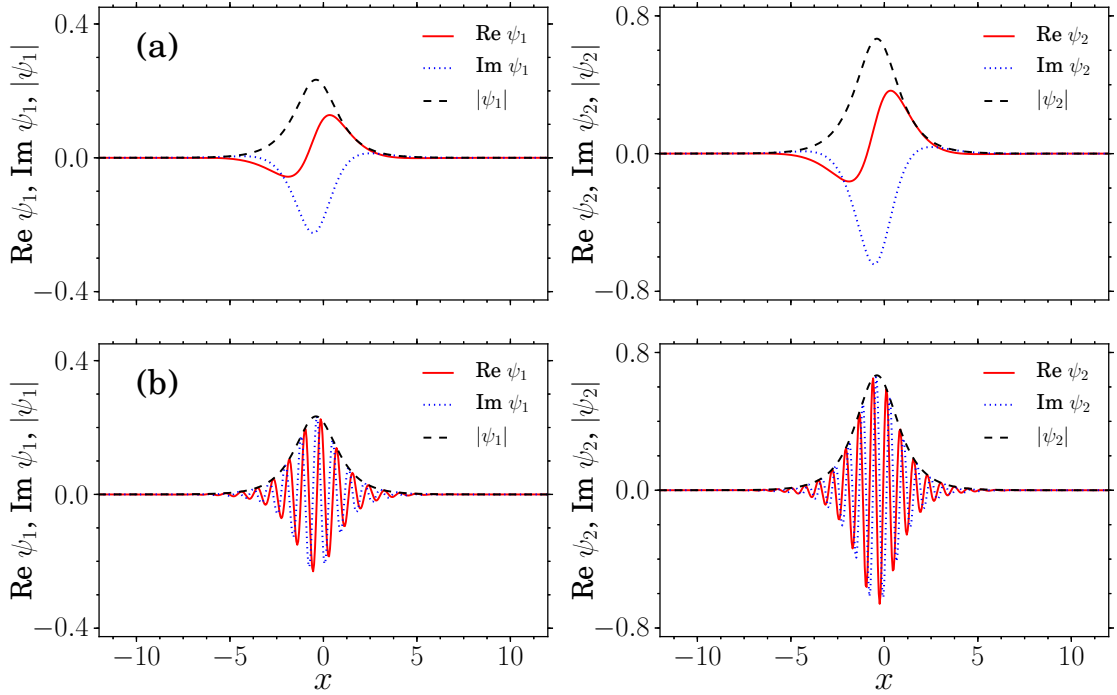}
\caption{Real, Imaginary parts and absolute value of the order
parameters $\psi_1$ and $\psi_2$ from (\ref{eq:sotrans}) and
(\ref{sol:q}) at time $t=0$ with $c(t)=0.04t$: (a) without
spin-orbit coupling, $k_L=0$ and (b) with spin-orbit coupling
$k_L=8$. The other parameters are $\alpha_{0} = 0.31$, $\beta_{0}
= 0.5$, $\gamma_0 = 2$, $\chi_1 = 0.5$, $\delta_1 = 0.2$,
$\varepsilon_{1} = 0.33$ and $\varepsilon_{2} =
\sqrt{1-\varepsilon_{1}^2}$. } \label{fig2-soc}
\end{figure}

Choosing a transient trap shown in figure~\ref{Fig.1.trap}, we show
the real, imaginary and absolute value of the macroscopic order
parameters $\psi_1$ and $\psi_2$ without and with spin orbit
coupling in figures~\ref{fig2-soc}(a) and \ref{fig2-soc}(b).
Comparison of figures~\ref{fig2-soc}(a) and \ref{fig2-soc}(b)
indicates that the addition of SO coupling contributes to the
rapid oscillations of the real and imaginary parts of  order
parameters $\psi_1$ and $\psi_2$. %
\begin{figure}[!ht] %
\centering\includegraphics[width=0.9\linewidth]{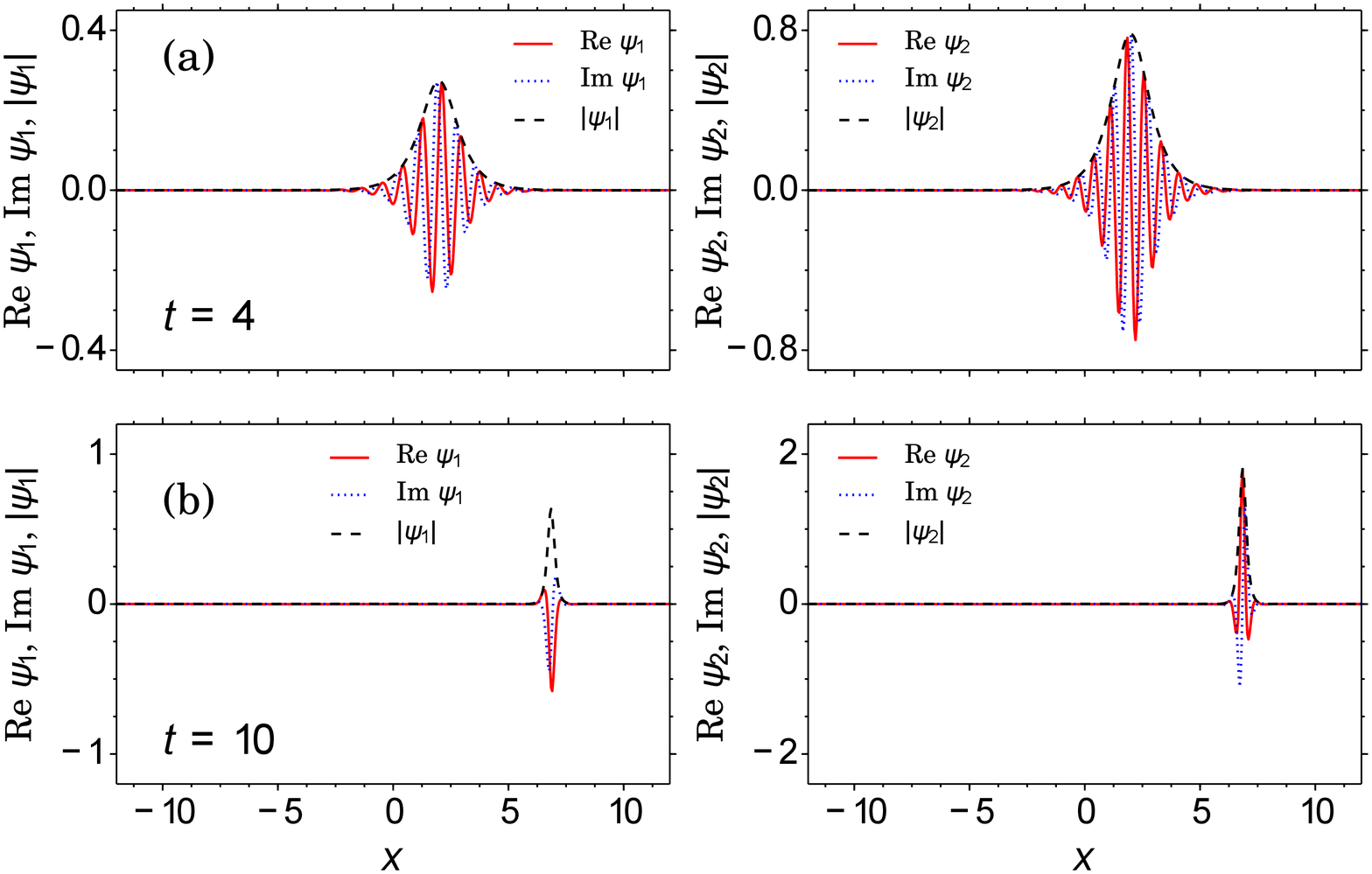}
\caption{Plots showing collapse of bright solitons in the
transient trap for $k_L=8$ and $c(t)=0.04t$ from
(\ref{eq:sotrans}) and (\ref{sol:q}) at (a) $t=4$ and (b) $t=10$
with the other parameters as in figure~\ref{fig2-soc} (see
figure~\ref{fig2-soc}(b) for $t=0$).}
\label{Fig.3.timedeptrapreim1}
\end{figure} %
The fact that the phase of $\psi_1$ and $\psi_2$ oscillates with
space and time as it is evident from (\ref{eq:sotrans})
contributes to oscillating nature of spin orbit coupled real and
imaginary parts of order parameter. Keeping track of the evolution
of the condensates in the transient trap, we observe that the
condensate collapses during time evaluation as shown in
figure~\ref{Fig.3.timedeptrapreim1}(b) for $t=10$. In other words,
the oscillating nature of real and imaginary parts of order
parameter is sustained as long as the trap remains confining in
nature ($\lambda(t)^2>0$) and the condensates collapse as soon as
the trap becomes expulsive ($\lambda(t)^2<0$). In addition to the
expulsive trap, the fact that the scattering length $\gamma(t)$ is
driven by $\exp (0.02 t^2)$ [by virtue of (\ref{integ})]
contributes to the collapse of BECs in quasi one dimension.
\begin{figure}[!ht]
\centering\includegraphics[width=0.9\linewidth]{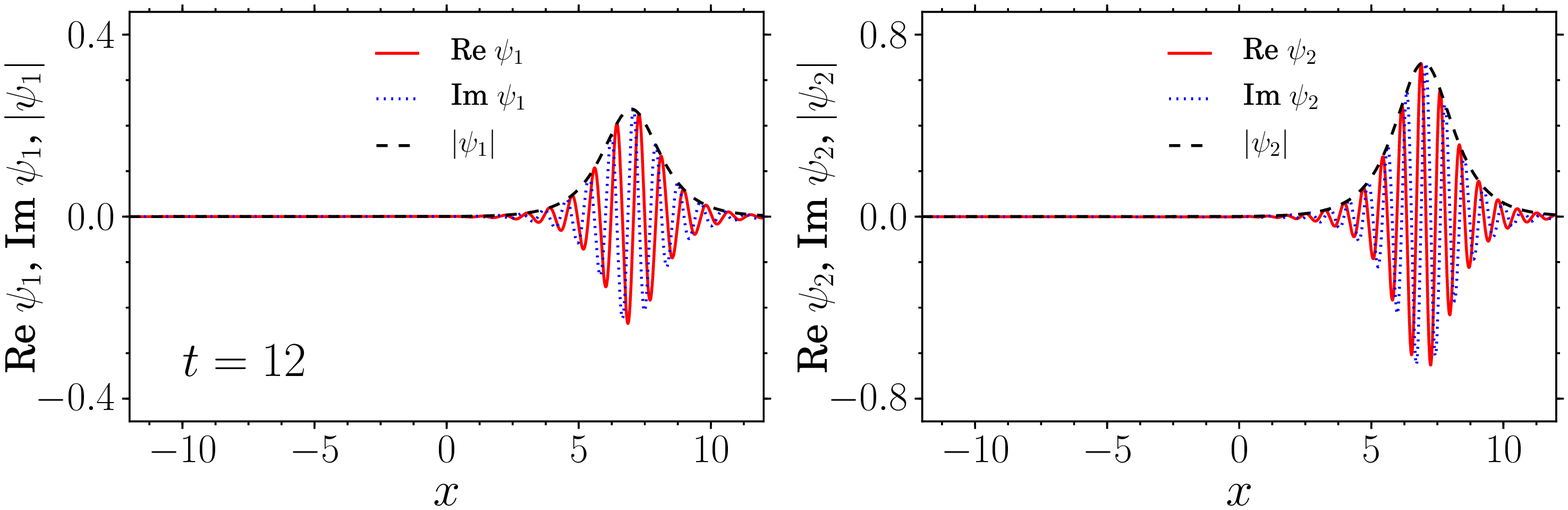}
\caption{Retrieval of bright solitons by stretching the confining
trap, that is, $c(t) = 0.0004t$ from (\ref{eq:sotrans}) and
(\ref{sol:q}) at $t=12$ (see figure~\ref{fig2-soc}(b) for $t=0$).
The other parameters are the same as in figure~\ref{fig2-soc}.}
\label{Fig.4.timedeptrapnew1reim}
\end{figure} %
However, by employing Feshbach resonance and manipulating the trap
frequency appropriately, one can increase the longevity of the
confining trap and recover the condensates as shown in
figure~\ref{Fig.4.timedeptrapnew1reim}. For a quasi one dimensional
attractive BEC confined in a harmonic trap with frequencies
$\omega_x= 2 \pi \times 20$Hz and $\omega_\perp= 2 \pi \times
1000$Hz, the trap frequency becomes
$({\omega_{x}}/{\omega_\perp})^2=0.0004$. The fact that the
condensates are recovered for $c(t) =0.0004 t$ (see
figure~\ref{Fig.4.timedeptrapnew1reim}) is consistent with the results
observed recently in~\cite{moto}.
\begin{figure}[!ht]
\centering\includegraphics[width=0.85\columnwidth]{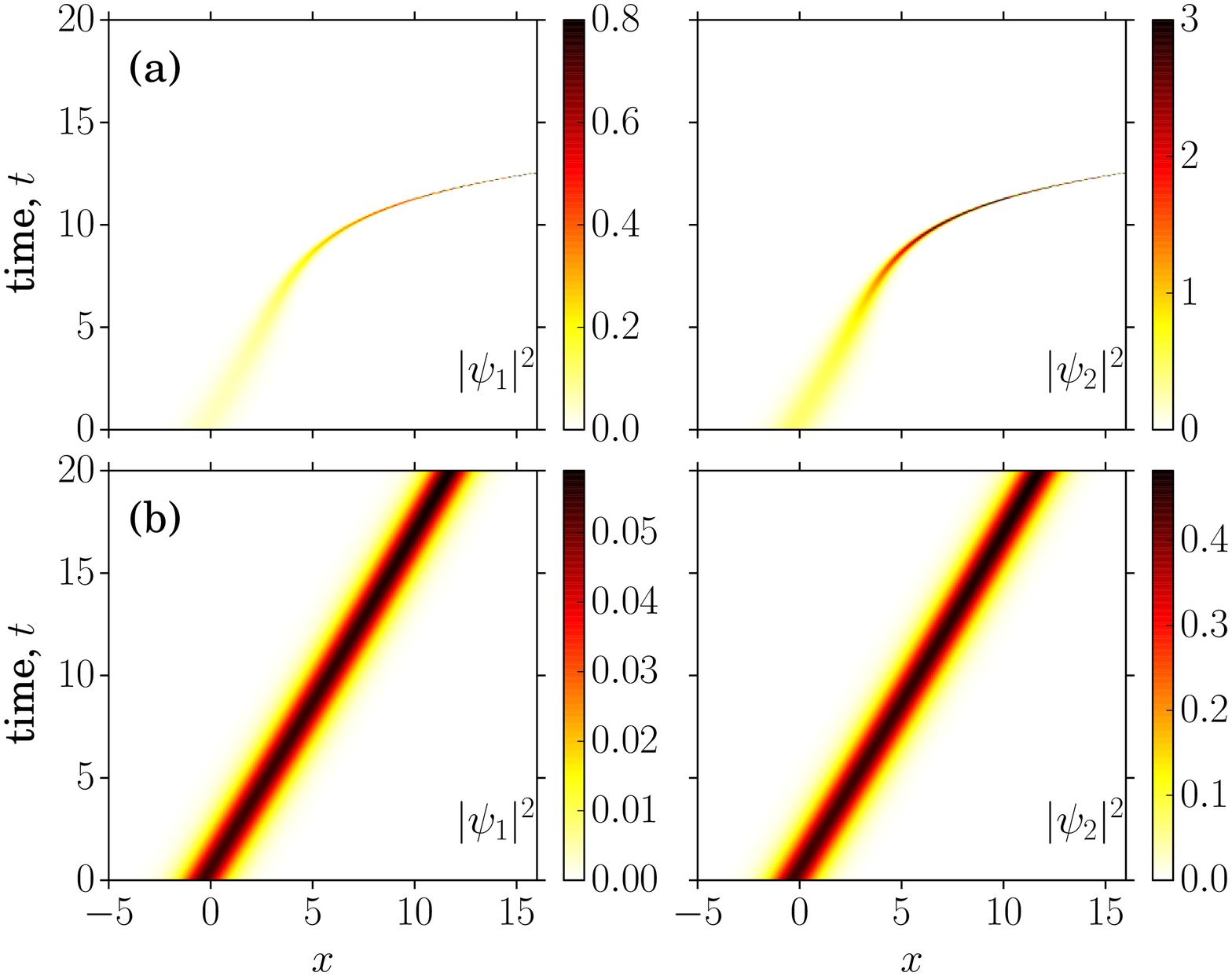}
\caption{Contours plots of densities $\lvert \psi_1 \rvert^2$ and
$\lvert \psi_2\rvert^2$ as a function  of time from
(\ref{eq:sotrans}) and (\ref{sol:q}) for (a) $c(t)=0.04t$ and (b)
$c(t)=0.0004t$ with $k_L = 8$ and the other parameters being the
same as in figure~\ref{fig2-soc}.} \label{Fig.5.numerical}
\end{figure}%
The stabilization of the condensates by manipulating the trapping
frequency through Feshbach resonance is also numerically verified
and the corresponding density profiles are shown in
figure~\ref{Fig.5.numerical}. It should also be emphasized that the
present integrable model offers the luxury of retrieving the
 condensates by tuning the time dependent
trap frequency through Feshbach resonance.
\begin{figure}[!ht] %
\centering\includegraphics[width=0.9\columnwidth]{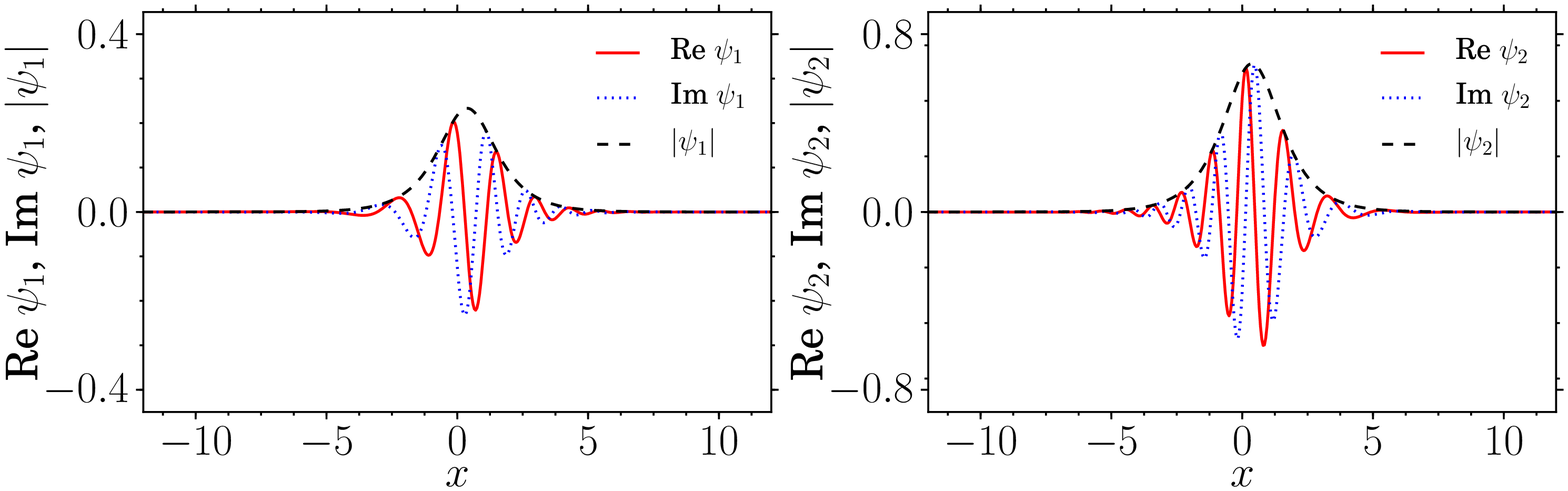}
\caption{Oscillating  real and imaginary  parts of the order
parameter in the time independent trap for a relatively small
choice of SO coupling $k_L=4$ and larger trap frequency
$c(t)=0.4$. The other parameters are the same as in
figure~\ref{fig2-soc}. }\label{Fig.6.timeinde}
\end{figure} %
Switching off the time dependence of the trap, one observes the
same oscillatory behavior for a relatively small  value of SO
coupling and larger trap frequency as shown in
figure~\ref{Fig.6.timeinde}. However, the subsequent time
evolution leads to collapse of the condensates which can once
again be stabilized employing Feshbach resonance.

We then study the impact of  Rabi coupling on bright solitons
numerically by considering the spin orbit coupled profile in
(\ref{eq:sotrans}) at $t=0$ as the initial condition.
\begin{figure}[!ht]
\centering\includegraphics[width=0.9\columnwidth]{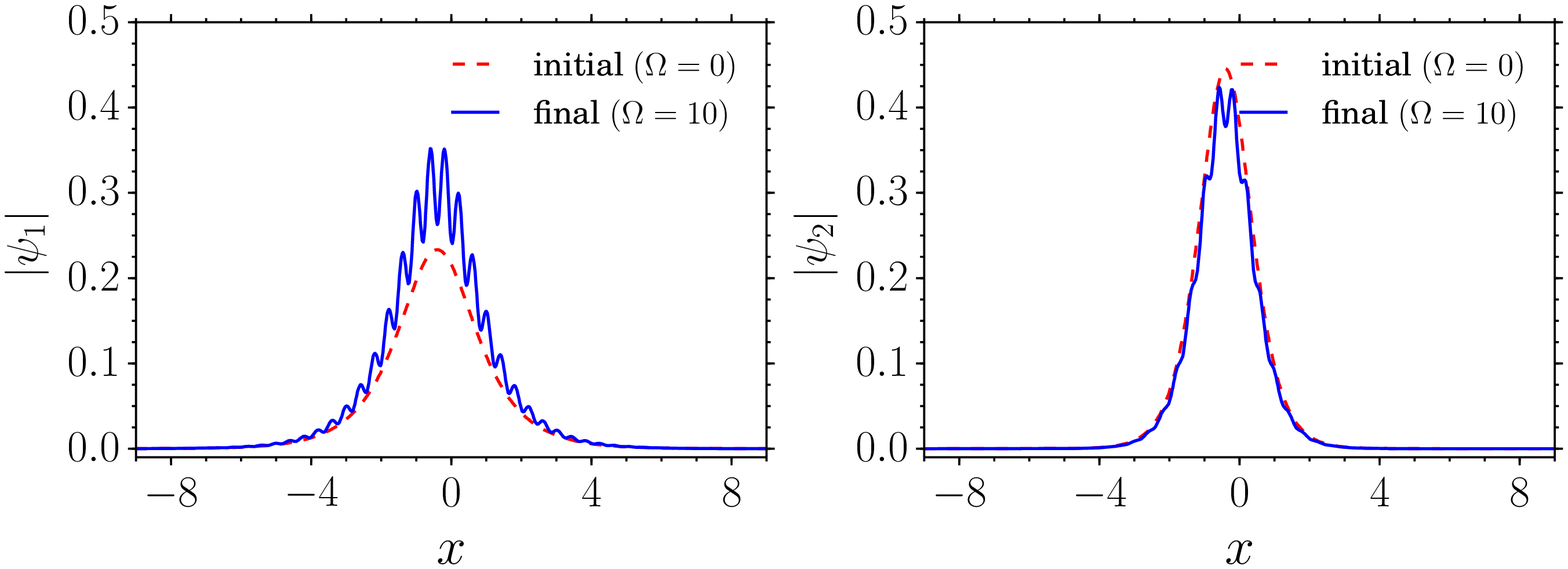}
\caption{Plots of the stationary states $\lvert \psi_1\rvert$ and
$\lvert \psi_2\rvert $ in the presence of Rabi coupling ($\Omega =
10$) with $k_L = 8$ and $c_0 = 0.0004$ as obtained from the
imaginary time propagation. The other parameters are the same as
in figure~\ref{fig2-soc}.} \label{Fig.numerical.1}
\end{figure} %
The numerical simulations are employed using a split-step
Crank-Nicolson
method~\cite{Muruganandam2009,Muruganandam20091,Muruganandam20092,Muruganandam20093}.
The initial profile is refined to be a stationary state using
imaginary time propagation, which is then evolved with realtime
propagation by adding the correct phase $\xi(x,0)$ and
$\theta(x,0)$ from (\ref{eq:xiphi}). Figure~\ref{Fig.numerical.1}
depicts the initial profile ($\Omega = 0$) from (\ref{eq:sotrans})
at $t=0$ and final profile of the stationary state as obtained
from imaginary time propagation with $\Omega = 10$.
\begin{figure}[!ht] %
\centering\includegraphics[width=0.85\columnwidth]{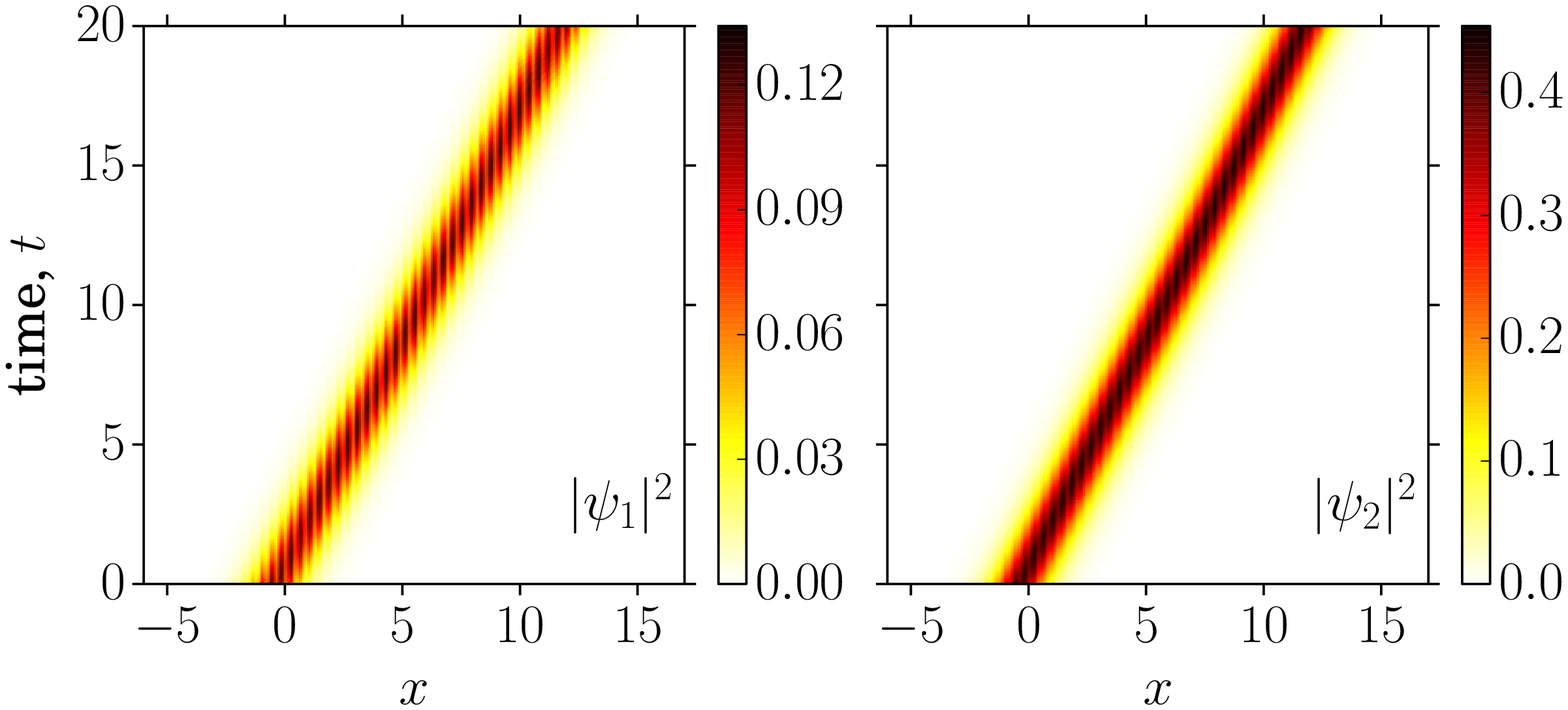}
\caption{Contour plots of densities $\lvert \psi_1\rvert^2$ and
$\lvert \psi_2\rvert^2$ as a function of time in the presence of
Rabi coupling ($\Omega = 10$) with $k_L = 8$ from the numerical
solution of (\ref{eq:gpe}) for $c(t)= c_0 = 0.0004t$. The other
parameters are the same as in figure~\ref{fig2-soc}.}
\label{Fig.7.numerical}
\end{figure} %
Figures~\ref{Fig.7.numerical} and \ref{Fig.8.numerical} show the
numerically simulated density profile of the SO coupled
condensates in the presence of Rabi coupling in a transient and
time independent harmonic trap respectively. The presence of
stripes in figures~\ref{Fig.7.numerical} and \ref{Fig.8.numerical}
is a clear signature of SO coupling consistent with the results of
\cite{moto}. The observation of striped phase which is an indication of the phase transition taking place in spin orbit coupled BECs is also consistent with the results of spin orbit coupled spin-1 BECs \cite{kueisun}.
Eventhough we have shown the system governed by (\ref{eq:gpe}) to
be completely integrable only in the absence of Rabi coupling
($\Omega=0$), we observe that the addition of Rabi coupling does
\begin{figure}[!ht]
\centering\includegraphics[width=0.85\columnwidth]{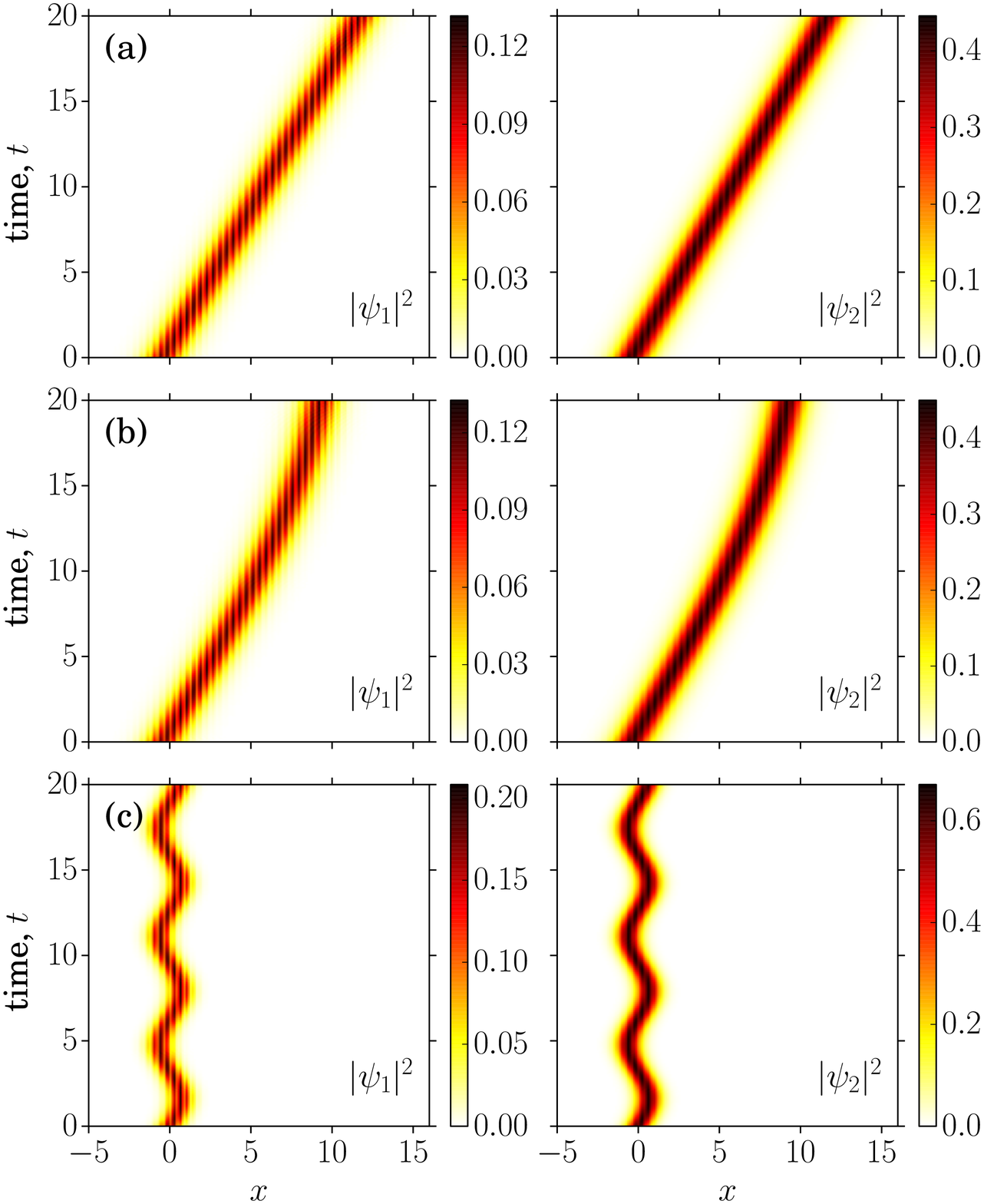}
\caption{Contour plots of densities $\lvert \psi_1\rvert^2$ and
$\lvert \psi_2\rvert^2$ as a function of time in the presence of
Rabi coupling ($\Omega = 10$) with $k_L = 8$ from the numerical
solution of (\ref{eq:gpe}) for different trap strengths  (a) $c_0
= 0.0004$, (b) $c_0 = 0.04$ and (c) $c_0 = 1$. The other
parameters are the same as in figure~\ref{fig2-soc}.}
\label{Fig.8.numerical}
\end{figure} %
not impact the stability of the condensates as it is evident from
 figures~\ref{Fig.7.numerical} and \ref{Fig.8.numerical}.
In other words, eventhough the addition of Rabi coupling to SO
coupled condensates takes the dynamical system to the
nonintegrable regime, one does not obviously observe its signature
on the condensates and they dwell in the stable regime. The
addition of different trap strengths (or different interaction
strengths) merely changes the trajectories of bright solitons
without impacting the stability of the condensates either as shown
in figures~\ref{Fig.8.numerical}(a)-(c). The tunability of transient harmonic trap to produce stable condensates is reminiscent of the tunability of stable SOC-BECs in a  double well  potential with the adjustable Raman frequency \cite{doublewell}.

\section{Conclusion}

In this paper, we have unearthed the Lax-pair of the SO coupled
BECs in a transient harmonic trap and shown them to be completely
integrable. We observe that the addition of SO coupling
contributes to the rapid oscillations in the real and imaginary
parts of order parameter. Tracing their evolution in a transient
harmonic trap, we observe that the condensates collapse in the
expulsive trap. However, by employing Feshbach resonance
management and manipulating trap frequency appropriately,  we are
able to retrieve the condensates. The oscillatory behavior in the
condensates occurs at a relatively small SO coupling parameter in
a time independent harmonic trap compared to its counterpart in a
transient trap. Then, considering a SO coupled state as the ground
state, we study numerically the impact of  Rabi coupling on BECs.
We observe that the reinforcement of  Rabi coupling in a SO
coupled BEC gives rise to striped solitons and one does not see
any signature of instability in the condensates. It would be
interesting to study the impact of SO coupling by considering a
Rabi coupled BEC as the initial state and the results will be
published later.

\section*{Acknowledgements}

RR wishes to acknowledge the financial assistance received from
Department of Atomic Energy-National Board for Higher Mathematics
(DAE-NBHM) (No. NBHM/R.P.16/2014) and Council of Scientific and
Industrial Research (CSIR) (No. 03(1323)/14/EMR-II) for the
financial support in the form Major Research Projects. The work of
PM forms a part of Science \& Engineering Research Board,
Department of Science \& Technology, Govt. of India sponsored
research project (No. EMR/2014/000644). PSV acknowledge the
support of UAE University through the grant UAEU-UPAR(7) and
UAEU-UPAR(4).

\end{document}